\documentclass[twocolumn,showpacs]{revtex4}   %%%%% actual paper style (for length estimate)

\usepackage{graphicx}

\begin{document}

%\preprint{}

\title{Physisorption of Nucleobases on Graphene}

\author{S. Gowtham$^1$}
\author{Ralph H. Scheicher$^{1,2}$}
\author{Rajeev Ahuja$^{2,3}$}
\author{Ravindra Pandey$^1$}\email[Corresponding authors. E-mail: ]{rhs@mtu.edu, pandey@mtu.edu}
\author{Shashi P. Karna$^4$}

\affiliation{$^1$Department of Physics and Multi-Scale Technologies
Institute, Michigan Technological University, Houghton, Michigan
49931, USA}

\affiliation{$^2$Condensed Matter Theory Group, Department of
Physics, Box 530, Uppsala University, S-751 21 Uppsala, Sweden}

\affiliation{$^3$Applied Materials Physics, Department of Materials
and Engineering, Royal Institute of Technology (KTH), S-100 44
Stockholm, Sweden}

\affiliation{$^4$US Army Research Laboratory, Weapons and Materials
Research Directorate, ATTN: AMSRD-ARL-WM; Aberdeen Proving Ground,
Maryland 21005-5069, USA}

\date{\today}

\begin{abstract}
We report the results of our first-principles investigation on the
interaction of the nucleobases adenine (A), cytosine (C), guanine
(G), thymine (T), and uracil (U) with graphene, carried out within
the density functional theory framework, with additional
calculations utilizing Hartree--Fock plus second--order
M{\o}ller--Plesset perturbation theory. The calculated binding
energy of the nucleobases shows the following hierarchy: G $>$ T
$\approx$ C $\approx$ A $>$ U, with the equilibrium configuration
being very similar for all five of them. Our results clearly
demonstrate that the nucleobases exhibit significantly different
interaction strengths when physisorbed on graphene. The stabilizing
factor in the interaction between the base molecule and graphene
sheet is dominated by the molecular polarizability that allows a
weakly attractive dispersion force to be induced between them. The
present study represents a significant step towards a
first-principles understanding of how the base sequence of DNA can
affect its interaction with carbon nanotubes, as observed
experimentally.
\end{abstract}


\maketitle

%\section{Introduction}

DNA-coated carbon nanotubes represent a hybrid system which unites
the biological regime and the nanomaterials world. They possess
features which make them attractive for a broad range of
applications, e.g., as an efficient method to separate carbon
nanotubes (CNTs) according to their electronic properties
\cite{Nakashima:2003, Zheng:2003a, Zheng:2003b}, as highly specific
nanosensors, or as an in vivo optical detector for ions. Potential
applications of single-stranded DNA (ssDNA) covered CNTs range from
electron sensing of various odors \cite{Johnson:2005}, to probing
conformational changes in DNA triggered by shifts in the surrounding
ionic concentration \cite{Strano:2006a}, and detection of
hybridization between complementary strands of DNA
\cite{Star:2006,Strano:2006b}. The interaction of DNA with CNT is
not limited to the outer surface of the tube; it has also been
experimentally demonstrated that ssDNA can be inserted into a CNT
\cite{Okada:2006}, further enhancing the potential applications of
this nano-bio system.

The details of the interaction of DNA with CNTs have not yet been
fully understood, though it is generally assumed to be mediated by
the $\pi$-electron networks of the base parts of DNA and the
graphene-like surface of CNTs. One would like to obtain a better
understanding of the binding mechanism, and the relative strength of
base-CNT binding as it is indicated experimentally from
sequence-dependent interactions of DNA with CNTs
\cite{Zheng:2003b,Johnson:2005}. In this Letter, we present the
results of our first-principles study of the interaction of
nucleobases with a graphene sheet as a significant step towards a
deeper understanding of the interaction of ssDNA with CNTs.

Previous theoretical studies focused on the adsorption of the
nucleobase adenine on graphite \cite{Ortmann:2005}. In the present
study, we have considered all five nucleobases of DNA and RNA,
namely the two purine bases adenine (A) and guanine (G), and the
three pyrimidine bases cytosine (C), thymine (T), and uracil (U).
Our specific interest is to assess the subtle differences in the
adsorption strength of these nucleobases on graphene, which in turn
will allow us to draw conclusions for the interaction of DNA and RNA
with CNTs as well.

%\section{Computational Method}

Calculations were performed using the plane-wave pseudopotential
approach within the local density approximation (LDA) \cite{LDA,
LDA-remark} of density functional theory (DFT) \cite{DFT}, as
implemented in the Vienna Ab-initio Simulation Package ({\sc vasp})
\cite{VASP}. The cutoff energy was set to 850 eV.  For $k$-point
sampling of the Brillouin zone we used the $1\times1\times1$
Monkhorst-Pack grid \cite{Monkhorst:1976}, which we found from
benchmark calculations to yield identical results as a
$3\times3\times1$ Monkhorst-Pack grid would.

A $5\times5$ array of the graphene unit cell in the $x$-$y$ plane
and a separation of 15 \AA\ between adjacent graphene sheets in the
$z$-direction was found to be a suitable choice to represent the
supercell. The base molecules were terminated at the cut bond to the
sugar ring with a methyl group in order to generate an electronic
environment in the nucleobase more closely resembling the situation
in DNA and RNA rather than that of just individual isolated bases by
themselves. This has the additional benefit that a small magnitude
of steric hindrance can be expected from the methyl group, quite
similar to the case in which a nucleobase with attached sugar and
phosphate group would interact with graphene.

For each of the five nucleobases, an ``initial force relaxation''
calculation step determined the preferred orientation and optimum
height of the planar base molecule relative to the graphene sheet. A
slice of the potential energy surface was then explored by
translating the relaxed base molecules in a fixed orientation
parallel to the graphene plane in steps of 0.246 \AA\ along the
lattice vectors of graphene, covering its entire unit cell by a mesh
of 10 $\times$ 10 scan points. The separation between base molecule
and graphene sheet was held fixed at the optimum height determined
previously. The determination of the minimum total energy
configuration was then followed by a 360$^\circ$ rotation of the
base molecules in steps of 5$^\circ$ to probe the dependence of the
energy on the orientation of the base molecules with respect to the
underlying 2-D graphene sheet. The configuration yielding the
minimum total energy was used in the final optimization step in
which all atoms in the system were free to relax. We would like to
emphasize here that for all five nucleobases, the eventually
determined equilibrium configuration was characterized by a
separation between base and graphene sheet that was equal to the
optimum height chosen in the previous lateral potential energy
surface scan.

An additional set of calculations was performed using the {\it ab
initio} Hartree--Fock approach coupled with second--order
M{\o}ller--Plesset perturbation theory (MP2) as implemented in the
{\sc gaussian 03} suite of programs \cite{GAUSSIAN}. Due to the use
of localized basis sets (rather than plane-wave), the system here
consisted of the five nucleobases on top of a patch of nanographene
\cite{Enoki:2000:2002}, i.e., a finite sheet containing 28 carbon
atoms. The LDA optimized configuration and the 6-311++G(d,p) basis
sets for C, H, N and O atoms were used for the MP2 calculations.

%\section{Results and discussion}

The first optimization step involving the ``initial force
relaxation'' led to a configuration of all five nucleobases in which
their planes are likewise oriented almost exactly parallel to the
graphene sheet with a separation of about 3.5 \AA, characteristic
for $\pi$--$\pi$ stacked systems \cite{pi-stacking}. The interaction
of the attached methyl group with the graphene sheet results in a
very small tilt of the molecule, with angles less than 5$^\circ$.

\begin{figure}[ht]
 \begin{center}
    \includegraphics[scale=0.25, angle=0.]{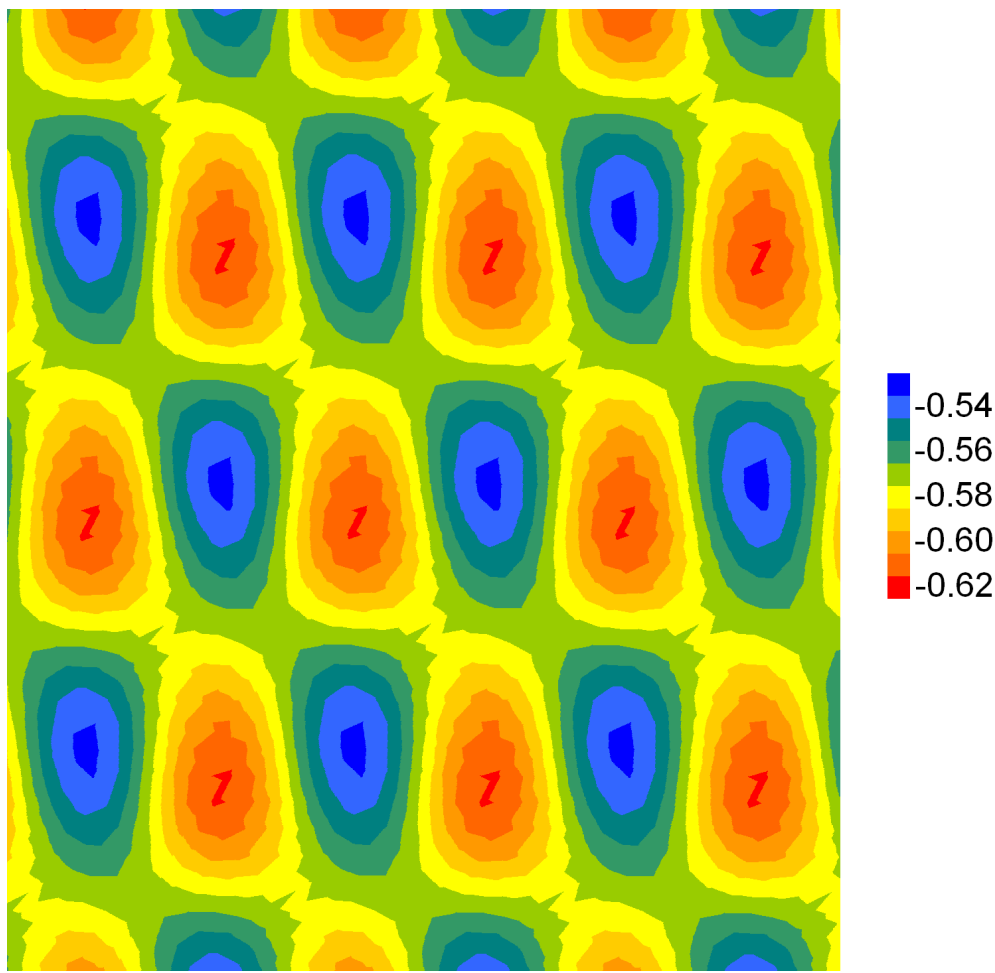}
    \caption{Potential energy surface (PES) plot (in eV) for guanine on
    graphene. Qualitatively similar PES plots were obtained
    for the other four nucleobases.
    Approximately a 3 $\times$ 3 repetition of the unit cell is shown.
    The energy range between peak and valley is approximately 0.1 eV.
    Energy barriers of only about 0.04 eV separate adjacent global minima.}
    \label{PES}
  \end{center}
\end{figure}

The base is translated 2.461 \AA\ along both graphene lattice unit
vectors respectively (maintaining a constant vertical distance of
3.5 \AA\ from the sheet, as determined in the previous step), and
rotated 360$^\circ$ in the equilibrium configuration with respect to
the configuration obtained after the ``initial force relaxation''
step in the optimization procedure. From the optimization steps
involving the translational scan of the energy surface, it is
apparent that the energy barriers to lateral movement of a given
base can range from 0.04 to 0.10 eV (Fig. \ref{PES}), thus
considerably affecting the mobility of the adsorbed nucleobases on
the graphene sheet at room temperature, and constricting their
movement to certain directions. The rotational scans carried out by
us found energy barriers of up to 0.10 eV, resulting in severe
hindrance of changes in the orientation of the adsorbed nucleobase.

\begin{figure}[ht]
 \begin{center}
    \includegraphics[scale=0.475, angle=0.]{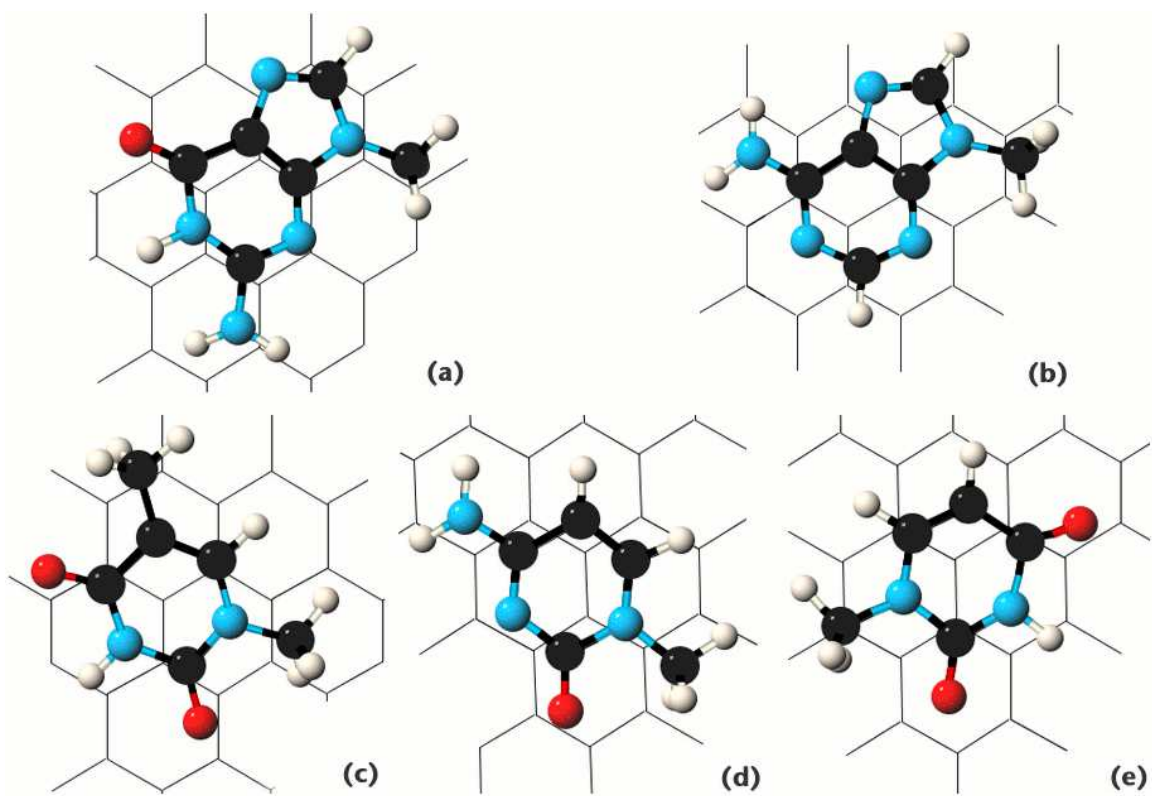}
    \caption{Equilibrium geometry of nucleobases on top of graphene: (a) guanine,
    (b) adenine, (c) thymine, (d) cytosine, and (e) uracil.}
    \label{equil_geom}
  \end{center}
\end{figure}

In their equilibrium configuration, three of the five bases tend to
position themselves on graphene in a configuration reminiscent of
the Bernal's AB stacking of two adjacent graphene layers in graphite
(Fig. \ref{equil_geom}). Virtually no changes in the interatomic
structure of the nucleobases were found in their equilibrium
configurations with respect to the corresponding gas-phase
geometries, as it could be expected for a weakly interacting system
such as the one studied here. A notable exception is the
R$_\mathrm{C-O}$ in guanine which shows a 10\% contraction upon
physisorption of the molecule on graphene.

The stacking arrangement shown in Fig. \ref{equil_geom} can be
understood from the tendency of the $\pi$--orbitals of the
nucleobases and graphene to minimize their overlap, in order to
lower the repulsive interaction. The geometry deviates from the
perfect AB base-stacking as, unlike graphene, the six- and
five-membered rings of the bases possess a heterogeneous electronic
structure due to the presence of both nitrogen and carbon in the
ring systems. In addition, there exist different side groups
containing CH$_3$, NH$_2$, or O, all of which contribute to the
deviation from the perfect AB base-stacking as well. Adenine,
thymine and uracil display the least deviation from AB stacking
(Fig. \ref{equil_geom}) out of the five nucleobases. For guanine and
cytosine on the other hand, there is almost no resemblance to the AB
stacking configuration recognizable (Fig. \ref{equil_geom}).

We calculated the binding energy for all five nucleobases. The
binding energy of the system consisting of the nucleobase and the
graphene sheet is taken as the energy of the equilibrium
configuration with reference to the asymptotic limit obtained by
varying the distance between the base and the graphene sheet in the
$z$-direction (Table \ref{BE_pol}). Within LDA, we found adenine,
cytosine and thymine to all possess nearly identical binding
energies of about 0.49 eV, while guanine with 0.61 eV is bound more
strongly, and uracil with 0.44 eV somewhat more weakly.

\begin{figure}[ht]
 \begin{center}
    \includegraphics[scale=0.5, angle=-90.]{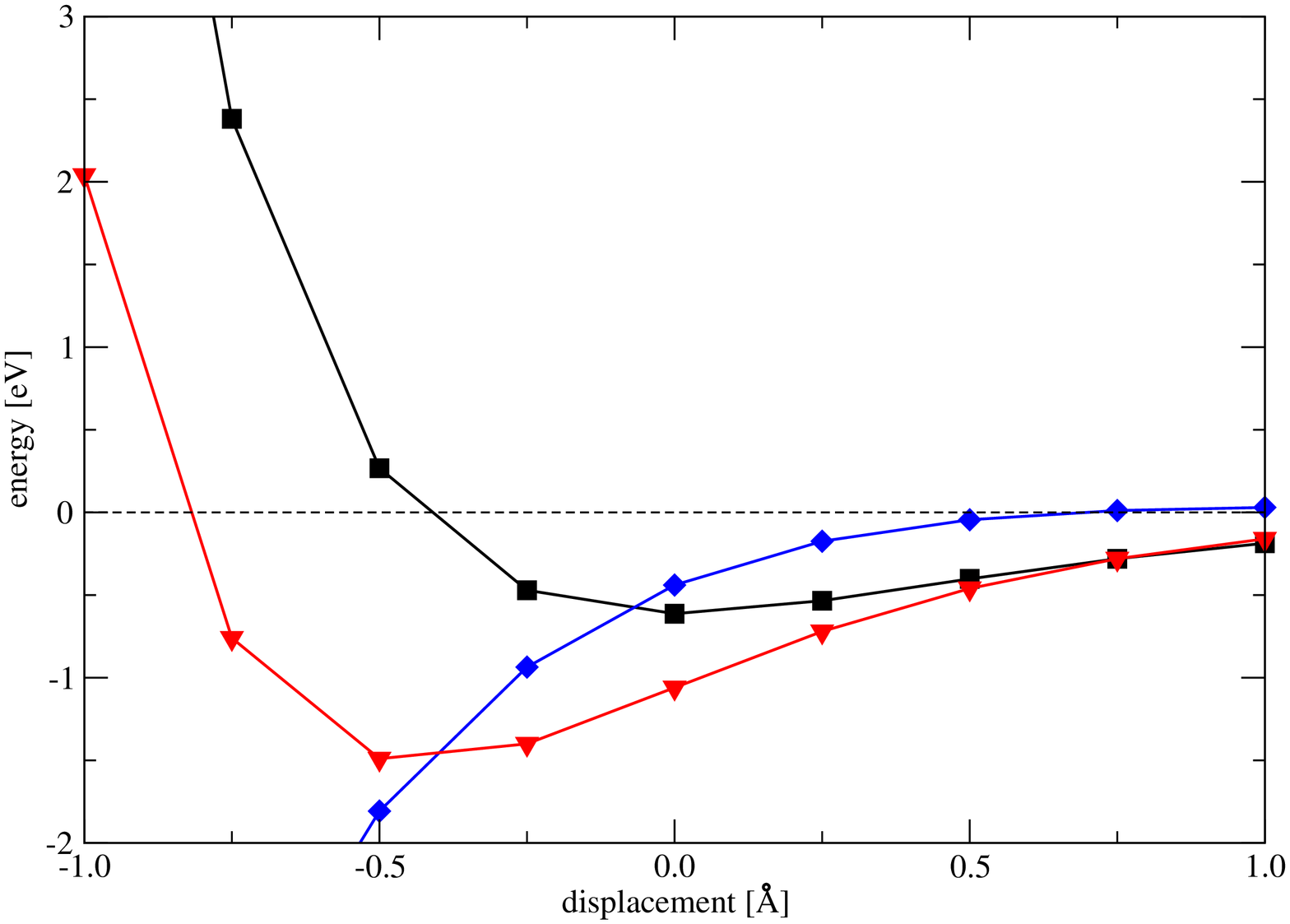}
    \includegraphics[scale=0.5, angle=-90.]{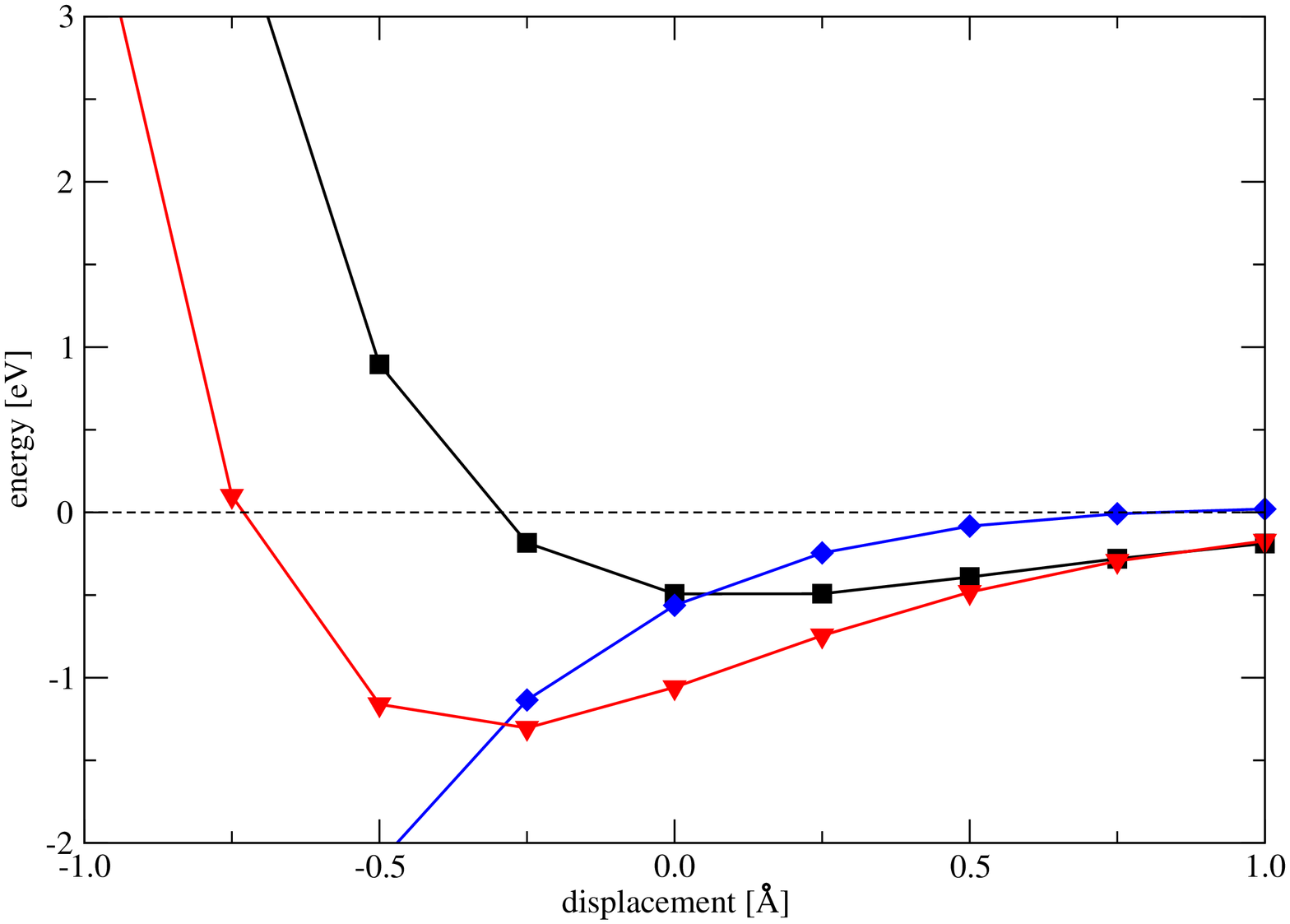}
    \caption{Plot of the relative total (black), exchange-correlation (blue)
and kinetic energy (red) of guanine (top) and adenine (bottom)
adsorbed on graphene calculated as a function of the displacement
from their respective equilibrium position.}
    \label{XC_kin_plots}
  \end{center}
\end{figure}

It is somewhat surprising that guanine and adenine would possess
such different physisorption energies, despite both containing a
five- and a six-membered ring and featuring relatively similar
molecular structures. A closer analysis of the various contributions
to the total energy (Fig. \ref{XC_kin_plots}) reveals that the
Kohn-Sham kinetic energy displays a slightly more pronounced minimum
for guanine than for adenine, and that the position of that minimum
is shifted by about 0.25 \AA\ towards the graphene sheet. The
exchange-correlation energy drops off somewhat more rapidly in the
case of adenine; however, the difference to the case for guanine is
only very small.

Table \ref{BE_pol} also includes the polarizabilities of the
nucleobases calculated at the MP2 level of theory. The
polarizability of the nucleobase \cite{Seela:2005}, which represents
the deformability of the electronic charge distribution, is known to
arise from the regions associated with the aromatic rings, lone
pairs of nitrogen and oxygen atoms. Accordingly, the purine base
guanine appears to have the largest value, whereas the pyrimidine
base uracil has the smallest value among the five nucleobases. Our
calculations confirm this behavior.

A remarkable correlation is found when the molecular
polarizabilities of the base molecules  are compared with the
binding energies, in particular when the latter are also determined
at the MP2 level of theory (Table \ref{BE_pol}). Clearly, the
polarizability of a nucleobase is the key factor which governs the
strength of interaction with the graphene sheet. This behavior is
expected for a system that draws its stabilization from van der
Waals (vdW) dispersion forces, since the vdW energy is proportional
to the polarizabilities of the interacting entities. The observed
correlation thus strongly suggests that vdW interaction is indeed
the dominant source of attraction between graphene and the
nucleobases.

The MP2 binding energies are systematically larger than those
calculated within the LDA approximation (Table \ref{BE_pol}). This
is due to the well established fact that MP2 provides a more
accurate treatment of the vdW interaction than LDA. We note that the
adsystem consisting of the base and the sheet is not bound at the
Hartree-Fock level of theory, which underscores the importance of
electron correlation in describing the weak vdW interactions in this
system.

In the equilibrium configuration, a redistribution of the total
charge density within a given base seems to appear. From an analysis
of the Mulliken charges for the MP2 calculations, we also find a
negligible charge transfer ($<$ 0.02 e) between any of the five
nucleobases and patch of nanographene in the equilibrium
configuration. Electrostatic interactions in the adsystem are
therefore very unlikely to contribute to the interaction energy.

%\section{Conclusions}

In summary, we investigated the physisorption of the five DNA/RNA
nucleobases on a planar sheet of graphene. Our first-principles
results clearly demonstrate that the nucleobases exhibit
significantly different interaction strengths when physisorbed on
graphene. This finding represents an important step towards a better
understanding of experimentally observed sequence-dependent
interaction of DNA with CNTs \cite{Zheng:2003b,Johnson:2005}. The
calculated trend in the binding energies strongly suggests that the
polarizability of the base molecules determines the interaction
strength of the nucleobases with graphene. As graphene can be
regarded as a model system for CNTs with very small surface
curvature, our conclusions should therefore also hold for the
physisorption of nucleobases on large-diameter CNTs. Further studies
involving the investigation of nucleobases interacting with
small-diameter CNTs are currently underway.

%\section*{Acknowledgments}

The authors acknowledge helpful discussions with Prof. Roberto
Orlando of the University of Turin, Italy, and with Dr. Takeru Okada
and Prof. Rikizo Hatakeyama of Tohoku University, Japan. S.G.,
R.H.S., and R.P. would like to thank DARPA for funding. R.H.S. and
R.A. are grateful to the Swedish National Infrastructure (SNIC) for
computing time. R.H.S. acknowledges support from EXC!TiNG (EU
Research and Training Network) under contract HPRN-CT-2002-00317.
The research reported in this document was performed in connection
with contract DAAD17-03-C-0115 with the U.S. Army Research
Laboratory.

\newpage

\begin{table}
\begin{center}
\begin{tabular}{cccc}
\hline \hline \multicolumn{1}{c} {base} & \multicolumn{1}{c}
{~~$E_b$(LDA) [eV]} & \multicolumn{1}{c} {~~$E_b$(MP2) [eV]} &
\multicolumn{1}{c} {~~$\alpha$ [$e^2a_0^2E_h^{-1}$]} \\
\hline
G  & 0.61  & 1.07 & 131.2 \\
A  & 0.49  & 0.94 & 123.7 \\
T  & 0.49  & 0.83 & 111.4 \\
C  & 0.49  & 0.80 & 108.5 \\
U  & 0.44  & 0.74 & 97.6 \\
\hline \hline
\end{tabular}
\end{center}
\caption{Binding energy $E_b$ of the DNA/RNA nucleobases with
graphene as calculated within LDA are compared with binding energy
and polarizability $\alpha$ from MP2 calculations.} \label{BE_pol}
\end{table}

\newpage

\newpage

\newpage

\begin{thebibliography}{99}

\bibitem{Nakashima:2003} N. Nakashima {\it et al.}, Chem. Lett. {\bf 32}, 456 (2003).

\bibitem{Zheng:2003a} M. Zheng {\it et al.}, Nature Mater. {\bf 2}, 338 (2003).

\bibitem{Zheng:2003b} M. Zheng {\it et al.}, Science {\bf 302}, 1545 (2003).

\bibitem{Johnson:2005} C. Staii {\it et al.}, Nano Lett. {\bf 5}, 1774 (2005).

\bibitem{Strano:2006a} D. A. Heller {\it et al.}, Science {\bf 311}, 508 (2006).

\bibitem{Star:2006} A. Star {\it et al.}, Proc. Natl. Acad. Sci. U.S.A. {\bf 103}, 921 (2006).

\bibitem{Strano:2006b} E. S. Jeng {\it et al.}, Nano Lett. {\bf 6}, 371 (2006).

\bibitem{Okada:2006} T. Okada {\it et al.}, Chem. Phys. Lett. {\bf
417}, 288 (2006).

\bibitem{Ortmann:2005} F. Ortmann, W. G. Schmidt, and F. Bechstedt,
Phys. Rev. Lett. {\bf 95}, 186101 (2005).

\bibitem{LDA} J. P. Perdew and A. Zunger, Phys. Rev. B {\bf 23}, 5048 (1981).

\bibitem{LDA-remark} LDA appears to give a reliable description of
dispersive interactions, unlike the generalized gradient
approximation (GGA) \cite{GGA} for which binding is basically
non-existent for van der Waals bound systems \cite{Simeoni:2005,
Tournus:2005}. The adsorption of adenine on graphite was recently
investigated \cite{Ortmann:2005} using a modified version of the
London dispersion formula \cite{London:1930} in combination with
GGA. The results, however, clearly indicate that LDA, while
underbinding the system, does in fact yield a potential energy
surface for adenine on graphite which is almost indistinguishable in
its structure from the one obtained via the GGA+vdW approach (cf.
Figs. 1a and 1b of Ref. \cite{Ortmann:2005}). LDA yields almost the
same equilibrium distance of adenine to graphene as GGA+vdW. The
source of the attraction is identified as the exchange-correlation
energy, either calculated within LDA or calculated within GGA+vdW.

\bibitem{GGA} J. P. Perdew {\it et al.}, Phys. Rev. B {\bf 46}, 6671 (1992).

\bibitem{Simeoni:2005} M. Simeoni {\it et al.}, J. Chem. Phys. {\bf
122}, 214710 (2005).

\bibitem{Tournus:2005} F. Tournus, S. Latil, M. I. Heggie, and J. C. Charlier, Phys. Rev. B. {\bf 72}, 075431 (2005).

\bibitem{London:1930} F. London, Z. Phys. {\bf 63}, 245 (1930); Z. Phys. Chem., Abt.
B {\bf 11}, 222 (1930).

\bibitem{DFT} P. Hohenberg and W. Kohn, Phys. Rev. {\bf 136}, B864 (1964); W.
Kohn and L. J. Sham, {\it ibid.} {\bf 140}, A1133 (1965).

\bibitem{VASP} G. Kresse and J. Furthm\"uller, Comput. Mater. Sci. {\bf 6}, 15
(1996); G. Kresse and D. Joubert, Phys. Rev. B {\bf 59}, 1758
(1999).

\bibitem{Monkhorst:1976} H. J. Monkhorst and J. D. Pack, Phys. Rev. B {\bf 13}, 5188 (1976).

\bibitem{GAUSSIAN} Gaussian 03, Revision C.02, M. J. Frisch {\it et al.}, Gaussian, Inc., Wallingford CT (2004).

\bibitem{Enoki:2000:2002} The dangling bonds at the edge of the
nanographene patch have been saturated by hydrogen atoms. For this
and other types of nanographene, see Y. Shibayama, H. Sato, T.
Enoki, and M. Endo, Phys. Rev. Lett. {\bf 84}, 1744 (2000), and K.
Harigaya and T. Enoki, Chem. Phys. Lett. {\bf 351}, 128 (2002).

\bibitem{pi-stacking} Klein, Cornelis and Cornelius S. Hurlbut, Jr., Manual of Mineralogy: after Dana, 20th
ed. (1985); Watson JD, Baker TA, Bell SP, Gann A, Levine M, Losick
R., Molecular Biology of the Gene. 5th ed. Pearson Benjamin
Cummings: CSHL Press (2004).

\bibitem{Seela:2005} Frank Seela, Anup M. Jawalekar, and Ingo M\"unster, Helevetica Chimica Acta {\bf 88}, 751 (2005).

\end{thebibliography}
\end{document}